\documentclass{custom}

\begin{document}

\title{Search for new physics with ATLAS at the LHC}

\author{V.A.~Mitsou}

\address{CERN, EP Division, CH-1211 Geneva 23, Switzerland\\
  and\\
  University of Athens, Physics Department, Nuclear and Particle Physics 
  Section, Panepistimioupolis, GR-157~71 Athens, Greece\\
  E-mail: Vasiliki.Mitsou@cern.ch}
  
\address{\rm On behalf of the ATLAS Collaboration}

\maketitle

\abstracts{
Due to the high energy and luminosity of the LHC, the ATLAS experiment has a
huge discovery potential for new physics. A Standard Model Higgs boson can be
discovered over the full range of allowed masses, and its mass should be
measured with a precision of about 0.1\%. The Higgs sector of the MSSM should be
fully explored by searches for supersymmetric Higgs bosons. Squarks and gluinos
can be discovered up to masses of 2.5~TeV and several precision measurements can
be performed in the SUSY sector. The existence of particles predicted by 
other theories beyond the Standard Model has been also investigated.}

\begin{minipage}{4.2truein}
  {\small Presented at the {\em Lake Louise Winter Institute 2000: From 
  Particles to the Universe}, Alberta, Canada, 20-26 February 2000.}
\end{minipage}

\section{Introduction}

ATLAS, a general purpose detector for proton-proton collisions, will be capable 
of exploring 
the new energy regime of 14~TeV which will become accessible at the Large Hadron
Collider (LHC). The LHC will be installed in the existing LEP tunnel at
CERN, and will run at a design luminosity of $10^{34}$~cm$^{-2}$s$^{-1}$. About 
30~fb$^{-1}$ are expected to be collected during the first three years, when the
machine will run at low luminosity ($10^{33}$~cm$^{-2}$s$^{-1}$), while about 
100~fb$^{-1}$ per year will be collected when running at design luminosity. An 
ultimate integrated luminosity of at least 300~fb$^{-1}$ is achievable. 
The discovery potential of ATLAS \cite{TDR} for new phenomena is discussed in 
the following.

\section{Higgs bosons}

The experimental observation of one or several Higgs bosons will be fundamental
for a better understanding of the mechanism of electroweak symmetry breaking. In
the Standard Model (SM), one doublet of scalar fields is assumed, leading to the
existence of one neutral scalar particle H. In the Minimal Supersymmetric
Standard Model (MSSM), on the other hand, at least two Higgs doublets are
required, corresponding to two charged (H$^{\pm}$) and three neutral (h, H, A) 
physical states. 

\subsection{Standard Model Higgs boson}

The dominant production mechanism of a SM Higgs boson at LHC energies is
gluon-gluon fusion, which proceeds via a heavy quark triangle loop. For larger
masses, also the WW fusion process contributes significantly. 

For Higgs masses below 150~GeV, the decay modes to ${\rm b\bar{b}}$
($BR\simeq90\%$) and to $\tau\tau$ ($BR\simeq10\%$) dominate. The decay to two 
photons is rather rare ($BR\sim10^{-3}$) and limited to the region 
$90<m_{\rm H}<150$~GeV. At larger masses ($>180$~GeV), the dominant decays are 
to WW ($BR\simeq75\%$) and to ZZ ($BR\simeq20\%$).

\begin{figure}[ht]
\centering
\epsfxsize=0.6\linewidth
\epsfbox{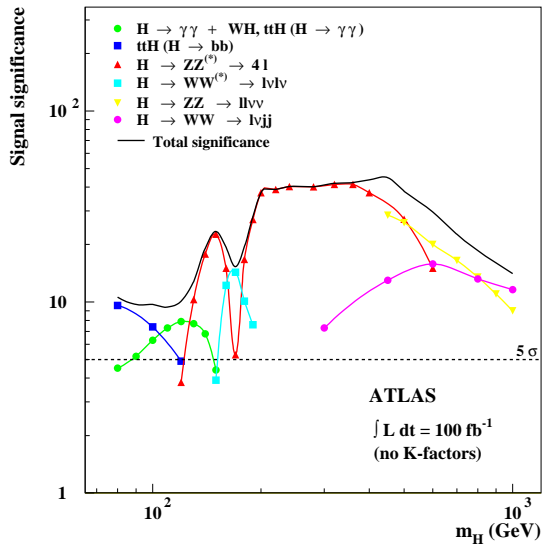} 
\caption{ATLAS sensitivity for the discovery of a Standard Model Higgs boson.
  The statistical significances are plotted as a function of the Higgs mass for 
  individual channels (different symbols), as well as for the combination of 
  all channels (full line), assuming an integrated luminosity of 100~fb$^{-1}$. 
  \label{fig:higgs}}
\end{figure}

The overall sensitivity for the discovery of a SM Higgs boson is shown in
Fig.~\ref{fig:higgs} for various channels assuming an integrated luminosity of
100~fb$^{-1}$. A SM Higgs boson can be discovered with the ATLAS experiment over
the full mass range up to $\sim$1~TeV with a high significance. A $5\sigma$
discovery can be achieved ---with two channels in most cases--- over the full 
mass range even after a few years of running at low luminosity. 

In the low mass region ($m_{\rm H}<150$~GeV), a SM Higgs boson can be discovered
through the ${\rm H}\rightarrow\gamma\gamma$ channel with a signal significance 
of 5--7$\sigma$ over the continuous $\gamma\gamma$ background.
The significance can be further enhanced by exploiting the associated production
of the Higgs boson with a W or a ${\rm t\bar{t}}$-pair.
In the same mass range, a signal from the ${\rm t\bar{t}H}$, 
${\rm H\rightarrow b\bar{b}}$ channel can also be observed with a significance 
$>5\sigma$ by exploiting the b-tagging capabilities of the detector.

The decay channel ${\rm H\rightarrow ZZ^*\rightarrow4\,\ell}$ provides a rather
clean signature in the mass range between $\sim$120~GeV and $2m_{\rm Z}$, above
which the gold-plated channel with two real Z bosons in the final states opens
up. Both electrons and muons are considered in the final state, thus yielding
eeee, ee$\mu\mu$ and $\mu\mu\mu\mu$ event topologies. 

In the mass region $150<m_{\rm H}<180$~GeV, a pronounced dip occurs in the 
$BR({\rm H\rightarrow ZZ^*})$, due to the opening of the WW decay mode. 
An excess of events from the 
${\rm H\rightarrow WW^{(*)}}\rightarrow\ell\nu\ell\nu$ channel can be used to 
enhance the significance in this region.

If a SM Higgs boson would be discovered at the LHC using the aforementioned
analyses, its mass will be measured with a precision of 0.1\% for 
$m_{\rm H}<400$~GeV and of 0.1--1\% for $400<m_{\rm H}<700$~GeV. 
The Higgs boson width can be determined for masses above 200~GeV using the
${\rm H\rightarrow ZZ\rightarrow4\,\ell}$ channel. 

\subsection{Minimal Supersymmetric Standard Model Higgs}

\begin{figure}[ht]
\centering
\epsfxsize=0.6\linewidth
\epsfbox{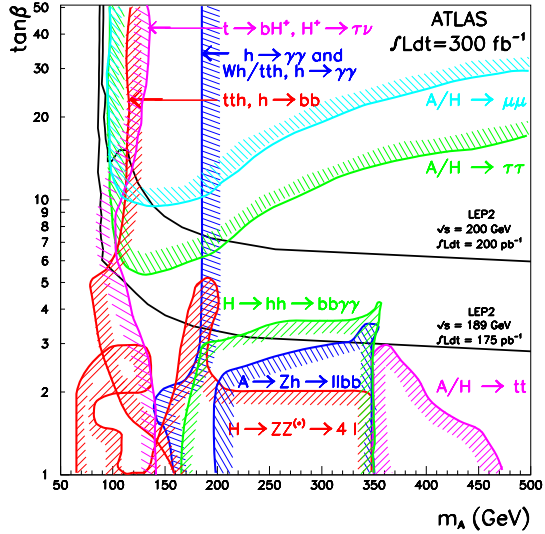} 
\caption{ATLAS sensitivity for the discovery of MSSM Higgs bosons. The $5\sigma$
  discovery contour curves for individual channels are shown in the 
  $(m_{\rm A},\,\tan\beta)$ plane for an integrated luminosity of 300~fb$^{-1}$.
  \label{fig:mssm}}
\end{figure}

The capability of the ATLAS experiment to detect MSSM Higgs bosons has been
studied in depth over the last few years. Large sparticles masses have been 
assumed so that Higgs bosons are not allowed to decay to supersymmetric 
particles. In the MSSM, various decay modes accessible also in the
case of the SM Higgs boson are predicted, such as 
${\rm h\rightarrow\gamma\gamma}$, ${\rm h\rightarrow b\bar{b}}$, 
${\rm H\rightarrow ZZ^{(*)}\rightarrow4\,\ell}$. In addition, some channels are 
strongly enhanced at large $\tan\beta$, e.g.\ ${\rm H/A\rightarrow\tau\tau}$ 
and ${\rm H/A\rightarrow\mu\mu}$. Other potentially interesting channels, such 
as ${\rm H/A\rightarrow t\bar{t}}$, ${\rm A\rightarrow Zh}$,
${\rm H\rightarrow hh}$ and ${\rm H^{\pm}\rightarrow tb}$, were also studied. 
The $5\sigma$ discovery contours for individual channels in the 
$(m_{\rm A},\,\tan\beta)$ plane are shown in Fig.~\ref{fig:mssm} for an 
integrated luminosity of 300~fb$^{-1}$. 

Complete coverage of the region shown will be possible at the LHC. Over a 
considerable fraction of the parameter space, at least two channels are 
accessible and/or more than one Higgs bosons can be observed. In most cases, 
the experiment will be capable of distinguishing between a SM and an MSSM Higgs 
boson.

\section{Supersymmetry}

\begin{figure}[ht]
\begin{minipage}[t]{0.47\linewidth}
\centering
\epsfxsize=0.92\linewidth
\epsfbox{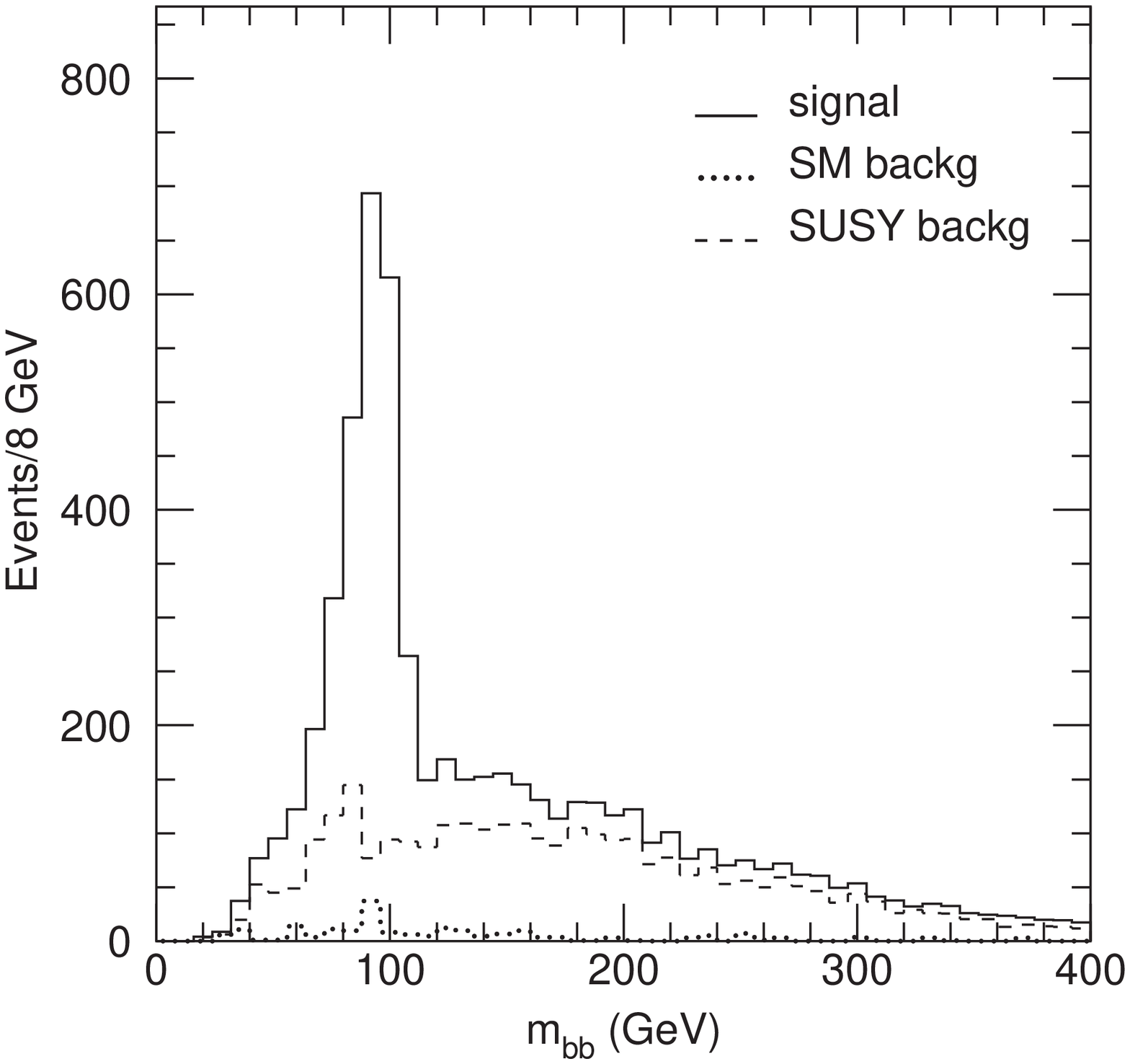} 
\caption{Mass distribution of two b-jets for SUGRA point: $m_0=100$~GeV, 
  $m_{1/2}=300$~GeV, $\tan\beta=2.1$, $A_0=300$~GeV and ${\rm sgn}\mu=+1$. The 
  ${\rm h\rightarrow b\bar{b}}$ signal (solid), the SUSY background (dashed) 
  and the SM background (dotted) are shown. \label{fig:mbb_5}}
\end{minipage}\hfill  
\begin{minipage}[t]{0.47\linewidth}
\centering
\epsfxsize=0.92\linewidth
\epsfbox{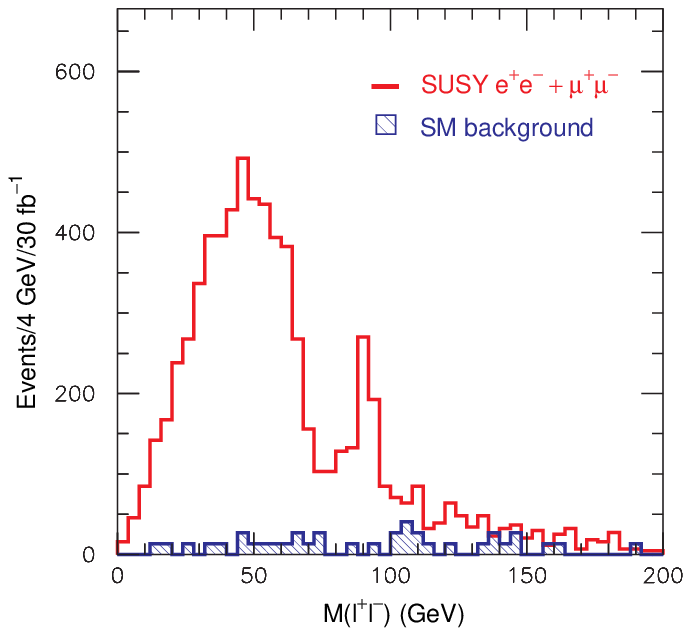} 
\caption{Dilepton distribution for SUGRA point: $m_0=800$~GeV, 
  $m_{1/2}=200$~GeV, $\tan\beta=10$, $A_0=0$ and ${\rm sgn}\mu=+1$. The SUSY 
  signal (solid) and the SM background (shaded) are shown. \label{fig:mll_4}}
\end{minipage}  
\end{figure}

If supersymmetry (SUSY) exists at the electroweak scale, then its discovery at 
the LHC should be straightforward. The SUSY cross section is dominated by 
gluinos and squarks production, which are strongly produced with large cross 
sections. Gluinos and squarks then decay via a series of
steps into the LSP (which may itself decay if $R$-parity is violated). These
decay chains lead to a variety of signatures involving multiple jets, leptons,
photons, heavy flavors (e.g.\ see Fig.~\ref{fig:mbb_5}), W and Z bosons, and 
missing energy. The combination of a
large production cross section and distinctive signatures makes it easy to
separate SUSY from the Standard Model background. Therefore, the main challenge
is not to discover SUSY, but to separate the many SUSY processes that occur and
to measure the masses and other properties of the SUSY particles. In most
cases, the backgrounds from other SUSY events dominate over the reducible SM
backgrounds. 

The approach followed has been the detailed investigation of signatures for 
particular points in the parameter spaces of the minimal supergravity (SUGRA), 
gauge mediated supersymmetry breaking (GMSB) and $R$-parity violating models. 
Methods such as looking for kinematic endpoints in mass distributions and using 
these to determine combination of masses have proven generally useful (e.g.\ 
see Fig.~\ref{fig:mll_4}). By using these methods, the fundamental parameters of
the underlying theory can be determined with precision of a few percent.

The starting point in the kind of analysis described earlier will be to look 
for characteristic deviations from the Standard Model. In SUGRA and some other 
models, there will be events with multiple jets and leptons plus large missing 
energy. In GMSB models, there would be events with prompt photons or 
quasi-stable sleptons. In $R$-parity violating models, there would be events 
with very high jet and/or leptons multiplicities. 

\section{Other physics scenarios}

Various theoretical scenarios, in addition to supersymmetry, have been studied 
by ATLAS in order to establish the discovery potential. ATLAS will be sensitive 
to new resonances predicted in technicolor theories, up to the TeV range. 
Although the parameter space is very large, the number of potential channels 
allows for combinations of signatures to help in understanding the nature of 
the resonances, and determine the possible existence of techniparticles.

Other studies involve the discovery of excited quarks in the photon plus jet
channel (masses up to 6~GeV), leptoquarks (masses up to 1.5~TeV) and 
compositeness probed by the high $p_{\rm T}$ jets (mass scale up to 40~TeV). 
New vector bosons can be discovered through their leptonic decays for masses up
to 5--6~TeV. Monopoles can be probed via the $\gamma\gamma{\rm X}$ cross section
for masses up to 20~TeV. It is also possible to investigate extra-dimensions
scenarios by searching for missing energy plus jet or missing energy plus photon
signatures.

\section{Conclusions}

ATLAS has a wide discovery potential for new physics and sensitivity to a
large variety of signatures. A Standard Model Higgs boson can be discovered, if
exists, over the whole mass range up to $\sim$1~TeV. MSSM Higgs bosons are
accessible over a large part of the $(m_{\rm A},\,\tan\beta)$ parameter space. 

Supersymmetric partners are expected to reveal themselves in signatures 
involving large missing energy, many leptons and/or many jets. These studies
have been performed in the context of
constrained supersymmetric scenarios, such as supergravity, gauge mediated
supersymmetry breaking and $R$-parity violation.

Moreover, the possibility of discovering new physics in many other cases has 
been investigated by ATLAS, such as technicolor, compositeness and extra
dimensions. Limits on the existence of excited quarks, leptoquarks, new gauge 
bosons, monopoles can be easily set up to the TeV scale. 


\section*{Acknowledgments}
I would like to thank the organizers of the Winter Institute for 
their hospitality and support during my stay at Lake Louise. 
This work was supported in part by the Greek State Scholarships Foundation 
(I.K.Y.).



\end{document}